\documentclass{aa} 

\usepackage{graphicx}
\usepackage{txfonts}
\usepackage{xcolor}
\usepackage{hyperref}

\def\Gyr{\,\mathrm{Gyr}}
\def\Myr{\,\mathrm{Myr}}
\def\kpc{\,\mathrm{kpc}}

\def\const{\mathrm{const}}
\def\kms{\,\mathrm{km\,s}^{-1}}

\def\Gaia{{\it Gaia}}

\begin{document}

   \title{The tangled warp of the Milky Way}

   \author{Viktor Hrannar Jónsson
          \inst{1}
          \and
          Paul J. McMillan \inst{2,1}
          }
   \institute{Lund Observatory, Division of Astrophysics, Department of Physics, Lund University, Box 43, SE-221\,00 Lund, Sweden 
         \and
         School of Physics \& Astronomy, University of Leicester, University Road, Leicester LE1 7RH, UK.\\
             \email{paul.mcmillan@leicester.ac.uk}
             }

   \date{Received XXX; accepted YYY}

 
  \abstract
   {
   }
   {We determine the influence of the Milky Way's warp on the kinematics of stars across the disc, and therefore measure its precession rate and line of nodes under different assumptions. }
   {We apply Jeans' first equation to a model of a rigidly precessing warp. The predictions of these models are fit to the average vertical velocities of stars with measured line-of-sight velocities in \Gaia\ DR3 data. We test models in which the warp's line of nodes and precession speed are fixed, and models in which they are allowed to vary linearly with radius. We also test models in which the velocity of stars radially in the disc is included in Jeans' equation.}
   {The kinematic data is best fit by models with a line of nodes that is $40^\circ$ offset from the Sun's Galactic azimuth, significantly leading the line of nodes found from the positions of stars. These models have a warp precession speed of around $13\kms\kpc^{-1}$ in the direction of Galactic rotation, close to other recent estimates. We find that including the velocity of stars radially in the disc in our kinematic model leads to a significantly worse fit to the data, and implausible warp parameters.}
   {The Milky Way's warp appears to be rapidly precessing, but the structure and kinematics of the warped disc are not consistent within the approximation of a fixed, precessing, warp shape. This implies that the Milky Way's warp is dynamically evolving, which is a challenge to models of the warp's creation, and must be considered in the context of other known disturbances of the disc.}

   \keywords{Galaxy: structure --
            Galaxy: kinematics and dynamics --
            Galaxy: disk --
            Galaxy: evolution 
               }

   \maketitle
%

\section{Introduction}
The Milky Way's disc, like that of most other spiral galaxies, is warped. The outer disc is bent upwards in roughly the direction that the Sun is currently moving, and downwards on the opposite side. This was first discovered in 21\,cm line measurements of the distribution of H\textsc{i} gas in the Galaxy \citep{kerr_mass_1957, burke_systematic_1957, westerhout_distribution_1957} before being found in the stellar population \citep{Djorgovski_iras_1989, Freudenreich_stellar_warp_1996} including the Cepheid population, where accurate distance estimation is possible to large distances \citep[e.g.,][]{chen_intuitive_2019, skowron_mapping_2019}. 
Similarly, warps have been found to be common features in both the gas \citep[e.g.,][]{bosma_21cm_1981} and stars \citep[e.g.,][]{sanchez-saavedra_frequency_1990} of spiral galaxies.

The warp influences the movement of stars and leaves a signature in Milky Way stellar kinematics. Systematic vertical velocity trends, dependent on Galactic positions or angular momentum, have long been attributed to the  warp 
(e.g., \citealt{dehnen_distribution_1998}; \citealt*{drimmel_galactic_2000}; \citealt{poggio_evidence_2020, cheng_exploring_2020}).
For a warp that was not precessing, we would simply expect a net upwards motion of stars moving towards the current higher end of the Milky Way's warp (and downwards for stars moving towards the lower end) in a predictable way. However, a warp that is precessing relatively quickly around the galaxy will, at a certain radius, begin to overtake the stars orbiting around the centre of the Galaxy, and therefore the sign of the stars' vertical velocity will reverse. This is a trend very clearly seen for the Milky Way's stars by \cite{cheng_exploring_2020}.

Most kinematic models of the warp assume that the velocity of stars towards or away from the centre of the Galaxy is irrelevant. This assumption underlies `tilted ring' models of galactic warps, which are commonly used for analysis of both external galaxies \citep*[e.g.,][]{rogstad_aperture-synthesis_1974,briggs_rules_1990} and the Milky Way \citep*[e.g.,][]{dehnen_a_twisted_2023}. It is also an assumption which is often put into models that apply Jeans' first equation to studying the Milky Way's warp \citep[e.g.,][]{drimmel_galactic_2000, poggio_evidence_2020}. However, a net radial velocity\footnote{Here, as throughout the paper, we use `radial' velocity to mean inwards or outwards in the disc. The component of velocity towards or away from the Sun we refer to as `line-of-sight' velocity} for a set of stars does correspond to a net movement within the warped disc, which one would expect to correspond to a net motion vertically. As shown by \cite{cheng_exploring_2020}, this enters naturally into Jeans' first equation and can be treated quantitatively in the corresponding models.

\begin{figure*}
    \centering
    \includegraphics[width=0.8\hsize]{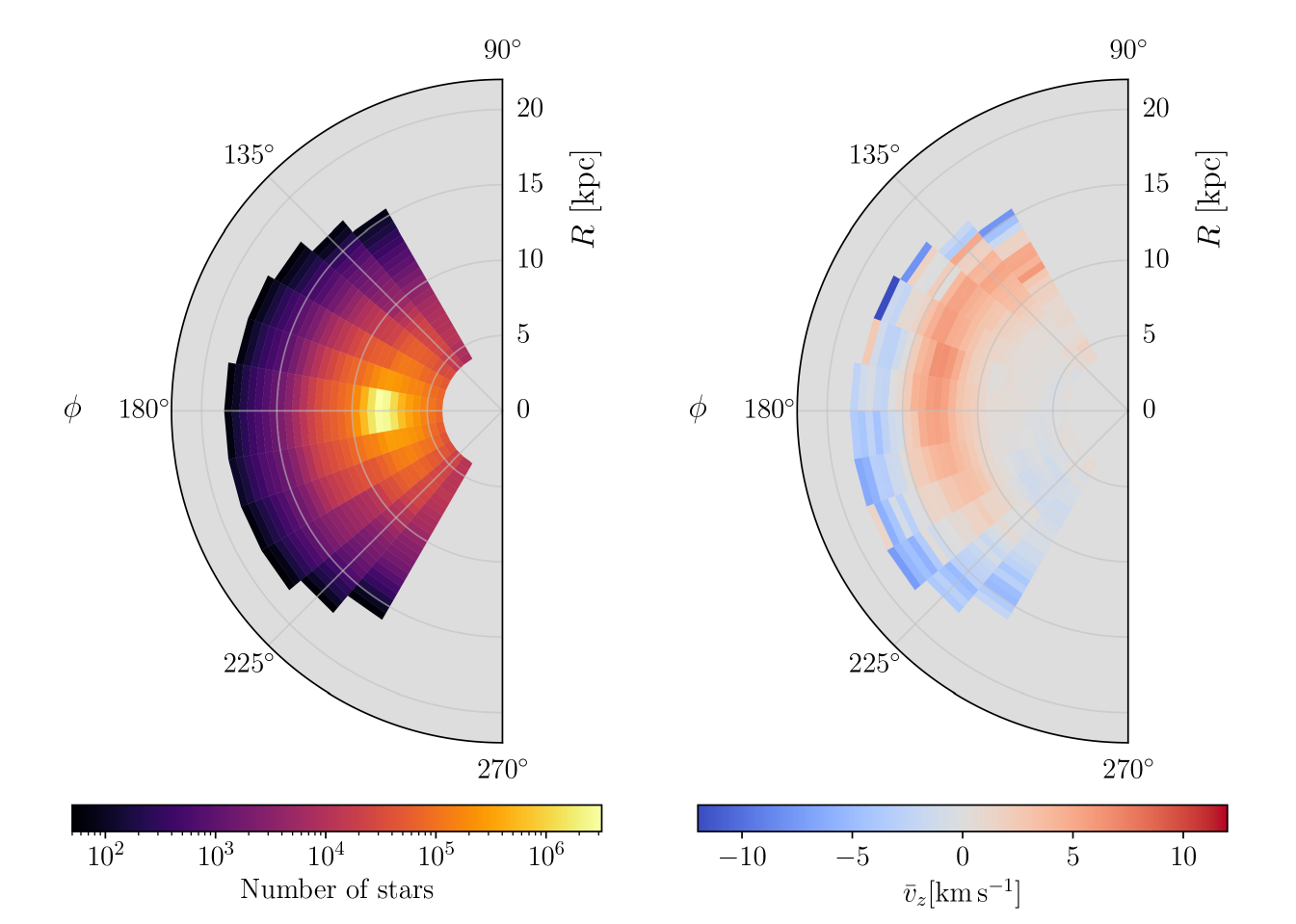}
    \caption{Distribution (left) and average vertical velocity (right) of our sample over the Galactic plane. This is binned, with each bin covering $0.5\kpc$ in $R$ and $10^\circ$ in $\phi$, both in galactocentric coordinates, with stars in the disc rotating in the negative $\phi$ direction.  Note that moving radially outwards from the position of the Sun $\Bar{v}_z$ rises to around 10 km s$^{-1}$ before falling to below zero, but that this behaviour disappears for $\phi\gtrsim220^\circ$}
    \label{fig:maps}
 \end{figure*}

 With \textit{Gaia} data release 2 \citep[DR2:][]{gaia_collaboration_gaia_2016, gaia_collaboration_summarydr2_2018} providing new insight into these stellar kinematics, the warp has been found to be precessing relatively rapidly in the prograde direction (the direction of the disc's rotation) at rates of 10-14 $\kms\kpc^{-1}$ \citep{poggio_galactic_2018, cheng_exploring_2020}. We note that  \cite{chrobakova_case_2021} dispute this finding, 
 though their estimate of $4^{+6}_{-4}\kms\kpc^{-1}$ is consistent with either a stationary or rapidly precessing warp. 
 For a sample of Cepheids from \textit{Gaia} data release 3 \citep[DR3:][]{gaia_collaboration_summary_2023} \cite{dehnen_a_twisted_2023} found a similarly high prograde precession rate of $9{-}15\kms\kpc^{-1}$. This is a much smaller number of stars, but with much more accurate and precise distance estimates, and can be considered an independent data sample and analysis. It also specifically probes the warp in the population of young stars, as Cepheids have ages of a few hundred million years or less. This precession is unexpected in the context of the standard picture of a warp made up of tilted rings of stars, in which the precession is expected to be retrograde and much slower. 
 

In most warp models, the line of nodes, $\phi_w$, along which the warped disc intersects $z=0$, is a parameter used to describe the orientation of the warp. A simple approximation that is suitable for relatively local samples, made for example by \cite{poggio_galactic_2018} and \cite{cheng_exploring_2020}, is that $\phi_w$ is a constant. However, \cite{briggs_rules_1990} found that the outer parts of warps in external galaxies are twisted, and that $\phi_w$ forms a leading spiral in the outer part of the warp (this is known as ``Briggs' rule''). 
For more extended samples the twist of the warp can be expected to be significant, and indeed both \cite{chen_intuitive_2019} and \cite{dehnen_a_twisted_2023} found that their Cepheid sample's warp followed Briggs' rule beyond $12\kpc$ from the Galactic centre, but the latter study also found that the warp precession rate decreased outwards such that it could unwind the twisted warp in ${\sim}100$ Myr.


The cause of the Milky Way's warp remains uncertain. Theories include the torque arisen from a misalignment between the Galactic disk and the dark matter halo \citep[e.g.,][]{dubinski_warps_2009}, accretion of intergalactic matter onto the disc \citep[e.g.,][]{ostriker_accretion_1989}, internal amplification of a small misalignment between inner and outer disc \citep{sellwood_internally_2021}, the interaction with the Large Magellanic Cloud \citep[e.g.,][]{kerr_magellanic_1957,burke_systematic_1957} or with the Sagittarius dwarf galaxy \citep[e.g.,][]{bailin_coupling_2003}. A better understanding of the warp's behaviour and kinematics is needed for us to compare the predictions of these theories to observations.

With the increased data set provided by \Gaia\ DR3, we are in a position to investigate the warp of the Milky Way over a large area, and to investigate the azimuthal variation of the warp's kinematic signature. We focus purely on the kinematics of the sample because the physical distribution of the sample seen in the \Gaia\ data is enormously affected by the dust extinction of the disc, which is not well understood. We therefore leave the study of the physical distribution of the warp to others, and will compare the results from our kinematic study to the physical warp found by these works. While doing this we are mindful that any limitations of the kinematic model will be translated into an alteration of the shape of the warp in the model. To keep a clear separation between results from kinematic models, and more direct measurements of the current physical shape of the warp, we use the subscript $_{phys}$ when describing the latter.

Our ability to trace the kinematics of the warp across a large area of the disc allows us to investigate its properties in great detail. It allows us to, for the first time, investigate the twist of the warp's line of nodes in the outer galaxy. We will explore the precession rate and line of nodes of the warp, and how they vary with radius. We will also examine the effect of including the radial velocity of stars in our kinematic models, following the pioneering study by \cite{cheng_exploring_2020}. These authors put all the stars in a given radial range in a single bin, which was characterised by its centre. The influence of the radial velocity on the model vertical velocity is zero along the line of nodes, and the centres of these bins tend to lay along the Sun's azimuth in the disc, which \cite{cheng_exploring_2020} had fixed as the line of nodes. Looking at the stars binned in azimuth gives us a much better lever arm to study the radial velocity's effect on the vertical velocity in this model, and we will show that this effect is significant.

In this paper we use \Gaia\ DR3 data to study the kinematic signature of the Milky Way's warp, and to measure its precession rate and line of nodes under different assumptions. Section~\ref{sec:data} describes the data sample taken from \Gaia\ DR3, and Sect.~\ref{sec:model} describes the kinematic models we use to fit these data. In Sect.~\ref{sec:fit_to_data} we show how the various models fit the measured average vertical velocities of stars across the Milky Way's disc, before looking at the properties of these models in Sect.~\ref{sec:properties}. Finally, we discuss these results in context and conclude the paper in Sects.~\ref{sec:discussion}~\&~\ref{sec:conclusions}, respectively.

\section{Data} \label{sec:data}
To study the kinematic signature of the Milky Way warp we need to know stars' three-dimensional positions and velocities, which \textit{Gaia} DR3 provides us to good accuracy for roughly 33 million stars. Since deriving the heliocentric distances from \textit{Gaia}'s parallaxes alone leads to biased and increasingly uncertain estimates at large distances, we use the \cite{bailer-jones_estimating_2021} photogeometric distances which combine \textit{Gaia} measured parallaxes with the colour and magnitudes of the star, and a Bayesian prior. The ADQL query used to obtain the data is given in Appendix~\ref{app:ADQL}.
To prevent the very worst distance errors from affecting our results we exclude stars for which the relative parallax error is large ($\varpi/\sigma_\varpi > 1$), and for which the \Gaia\ astrometric quality indicator RUWE $>1.4$ \citep{lindegren_astrometric_2021}.

We analyse the data in galactocentric cylindrical coordinates ($R, \phi, z$), where the position and velocity of the Sun with respect to the Galaxy's centre are taken to be the \texttt{astropy} default values\footnote{\href{https://docs.astropy.org/en/stable/coordinates/galactocentric.html}{Astropy galactocentric coordinates documentation}, `v4.0' parameters}. These are that the Sun is at a cylindrical galactocentric radius $R_\odot=8.122\kpc$, \citep{2018A&A...615L..15G}
and $z_\odot = 20.8$ pc above the Galactic plane,
with velocity $v_\odot = (v_{R,\odot}, v_{\phi,\odot}, v_{z,\odot}) = (-12.9, -245.6, 7.78)$ km s$^{-1}$ 
\citep{2018RNAAS...2..210D,2004ApJ...616..872R}.
The Sun's position in galactocentric azimuth is defined to be exactly $\phi_\odot = 180^\circ$. Since we use a standard right-handed coordinate system, the Milky Way disc's typical sense of rotation is negative, i.e. disc stars are rotating towards lower angles in azimuth. Where we cite studies that have used different conventions for the Sun's position in $\phi$, and for the direction of rotation, the values we quote in this paper will be converted into our coordinate system. 

We remove contaminants from the sample that come from the Large and the Small Magellanic Clouds (LMC and SMC), and 47 Tucanae. All stars within $12^\circ$ and $6^\circ$ respectively of the centres of the LMC and SMC respective on-sky positions\footnote{NASA/IPAC Extragalactic Database; \href{https://ned.ipac.caltech.edu/byname?objname=Large+Magellanic+Cloud&hconst=67.8&omegam=0.308&omegav=0.692&wmap=4&corr_z=1}{LMC}, \href{https://ned.ipac.caltech.edu/byname?objname=Small+Magellanic+Cloud&hconst=67.8&omegam=0.308&omegav=0.692&wmap=4&corr_z=1}{SMC}. (Accessed March 30th 2023)} were removed. For 47 Tucanae we remove stars in the range of $R \in (6.6, 6.9)\kpc$ and $\phi \in (200^\circ, 205^\circ)$ with $z < -2500\kpc$. 

The stars in our sample are divided up into bins in galactocentric radii and azimuth. In radii, the bins are 500 pc intervals beginning at $R=4\kpc$ and extending outwards to the edge of our sample, while the azimuth intervals are $10^\circ$ wide and extend from $\phi=120^\circ - 240^\circ$. We disregard any bin that contains fewer than 50 stars so that we can determine the average positions and velocities and their uncertainties from a sufficiently large sample.
This leaves us with a sample of $26\,891\,917$
stars in 327 bins. A map of the bins included in our sample, coloured by the number of stars in that bin, can be seen in Fig. \ref{fig:maps}. 

For each component of velocity we use the median as our average value, which we prefer to the mean as it is more robust to outliers and contamination. We derive uncertainties on the average of each velocity component $\Bar{v}_i$ using non-parametric bootstrapping. For each bin we resample the velocities 1000 times, and determine the median value in each case. Our statistical uncertainty $\sigma_{\Bar{v}_i, \text{stat}}$ is then the standard deviation of the distribution of resampled medians. 

As tests of the robustness of our results, we have look at data samples with a more conservative cut of 1000 stars per bin, and tests in which we consider stars within limited ranges of $R$. In this latter case we look at both $2\kpc$ annuli in $1\kpc$ steps with the outer $R$ ranging from $13$ to $16\kpc$ and at samples of stars with $R$ less than a threshold value which ranges from $13$ to $18\kpc$. We discuss the outputs of these tests when describing our results. 

The unprecedented number of stars in \textit{Gaia} DR3 for which we have full six dimensional phase space measurements allows us to probe the warp's kinematic signature as a function of azimuth, and Figure~\ref{fig:maps} also shows the average vertical velocity of the sample as a function of position in the Galaxy.  It is immediately obvious that the stars in the range $10\lesssim R\lesssim 15\kpc$ are moving upwards across most of the area covered, while those outside ${\sim}15\kpc$ are moving downwards. We can also see that beyond $\phi\approx220^\circ$ the stars in the range $10\lesssim R\lesssim 15\kpc$ are not moving upwards. As we will discuss through the rest of this paper, these two clear features of the data would appear to indicate that the warp is precessing quickly, and that the line of nodes of the warp, as defined by kinematics, lies closer to $\phi \approx 135^\circ$ than the value measured from the positions of stars of around $180^\circ$.

\section{The warp models} \label{sec:model}


\subsection{Warp shape}
The mean vertical position of the plane of the disc in our model, $z_0$, depends on both $R$ and $\phi$, and we separate these dependencies out, adopting
\begin{equation} \label{eq:z0_simple}
    z_0(R,\phi) = h(R) s(\phi, R)
\end{equation}
where $h(R)$ sets the greatest displacement due to the warp at a given $R$, while $s(\phi,R)$ is a function that takes values in the range $-1\leq s \leq 1$. 

For $h(R)$, numerical experiment led us to use a piecewise-linear model, with the free parameters being the peak heights at $1\kpc$ steps from $R=7\;\mathrm{to}\;16\kpc$. Inside $6\kpc$ we assume zero warp, while beyond $16\kpc$ we extrapolate $h(R)$ linearly from the last segment (or set it as constant if the final segment had a negative gradient -- our results do not depend on this approximation). We denote these $h_7,\;h_8,$..., $h_{16}$. For our robustness tests where there is a cut in $R$ we use the same $h(R)$ model, but limit it to the radii where the model affects the smaller dataset.

It has been common in studies of the Milky Way's warp to assume that the warp's dependence on azimuth, $s(\phi, R)$ in Eq.~\ref{eq:z0_simple}, is a simple sinusoidal function with no dependence on $R$  \citep[see][and others]{drimmel_galactic_2000, cheng_exploring_2020}. We follow this precedent and adopt the azimuthal dependence 
\begin{equation} \label{eq:s}
    s(\phi)=\sin(\phi - \phi'_w - \omega_pt),
\end{equation} 
where $\phi'_w$ is the line of nodes, and $\omega_p$ is the warp's precession rate. We present results using this model in this study. However, with the \textit{Gaia} DR3 data we can go further than this, and we therefore investigate the radial change (or `twist') of the line of nodes.

We therefore extend our model to allow both $\phi'_w$ and $\omega_p$ to vary linearly with $R$.  Specifically, we take
\begin{equation} \label{eq:s2}
    s(\phi, R)=\sin\left(\phi - \phi'_w(R) - \omega_p(R)\,t\right),
\end{equation} 
where
\begin{equation} \label{eq:wpLONpoly}
\begin{split}
    \phi'_w(R) &= \phi'_{w,0} + \phi_{w,1}(R-R_s) \\
    \omega_p(R) &= \omega_{p,0} + \omega_{p,1}(R-R_s), 
\end{split}
\end{equation}
where $R_s$ is a fixed at $12\kpc$. This is purely a numerical convenience that ensures that $\phi'_{w,0}$ and $\omega_{p,0}$ take values that are easy to interpret (i.e., the values of the respective parameters at $R=12\kpc$).
We note that, given our formulation of the problem, typical values of $\phi'_w(R)$ will correspond to the line of nodes on the far side of the Galactic disc. However, by symmetry the other line of nodes in our model is always at $\phi_w = \phi'_w + 180^\circ$, so it is this value that we quote, along with $ \phi_{w,0} =  \phi'_{w,0} + 180^\circ$ for models where the line of nodes varies with $R$.

\subsection{Modelling the kinematic signature}

Qualitatively, we can understand that stars moving from the lower part of the warp to the higher part will, today, be observed with a net upwards motion. For a warp that varies sinusoidally with $\phi$, this velocity will be a maximum at the line of nodes, and zero at the peak or trough of the warp. A first complication to this is that, for a precessing warp, the warp may be catching up with the stars. This can cause stars that appear to be heading towards the higher part of the warp to have a net downwards motion. A subtler effect is that stars moving inwards or outwards should also follow the warp shape, and this could also affect the expected velocities of the stars.

\begin{figure*}
    \centering    
    \includegraphics[width=0.9\linewidth]{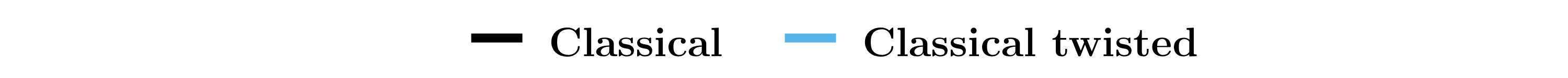}
    \includegraphics[width=0.9\linewidth]{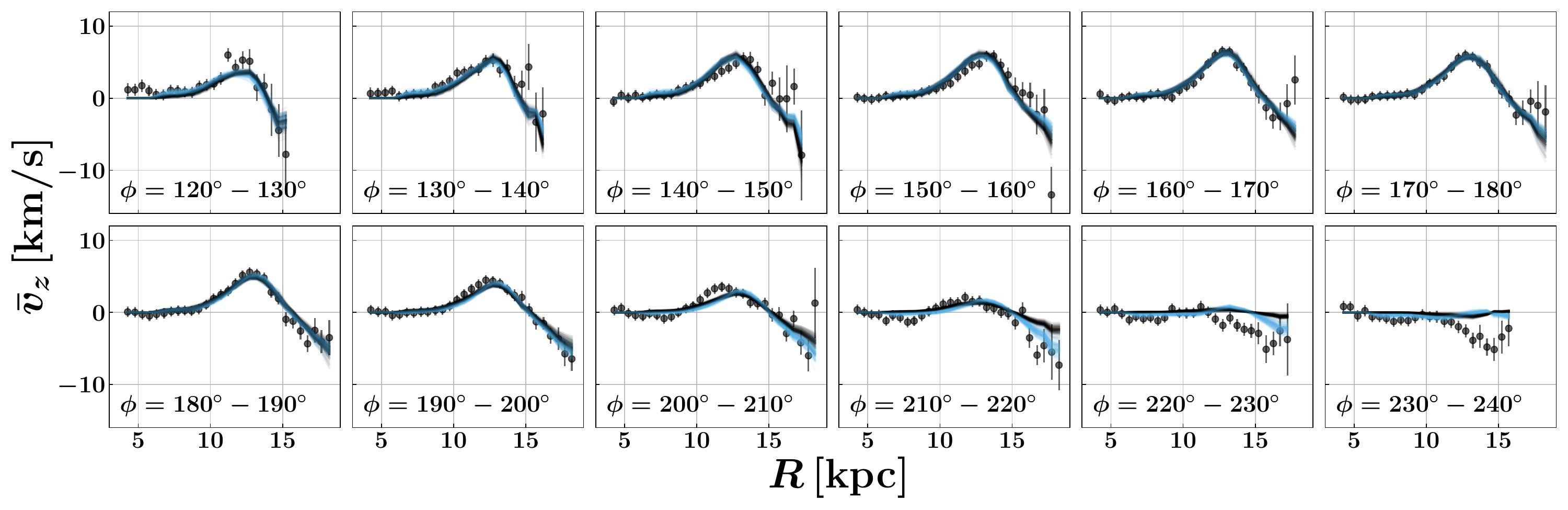}
    \caption{Fits to the $\Bar{v}_z$ data for \emph{Classical} models. Each panel shows data for $10^\circ$ sectors of the Galaxy, and plots the average $v_z$ in $0.5\kpc$ bins, with associated uncertainties (statistical and systematic) The model fits with a single value of $\phi_w$ and $\omega_p$ are plotted in black, while those of the \emph{twisted} model are in light blue. Overall both models provide reasonable fits, without major differences}
     \label{fig:Drimmel_fit}%
 \end{figure*}
 
 \begin{figure*}
    \centering
    \includegraphics[width=0.9\linewidth]{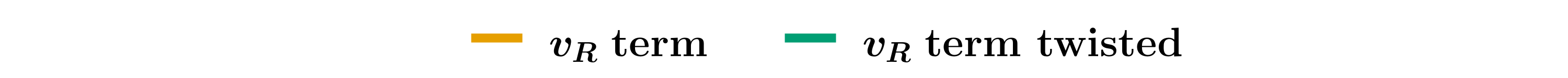}
    \includegraphics[width=0.9\linewidth]{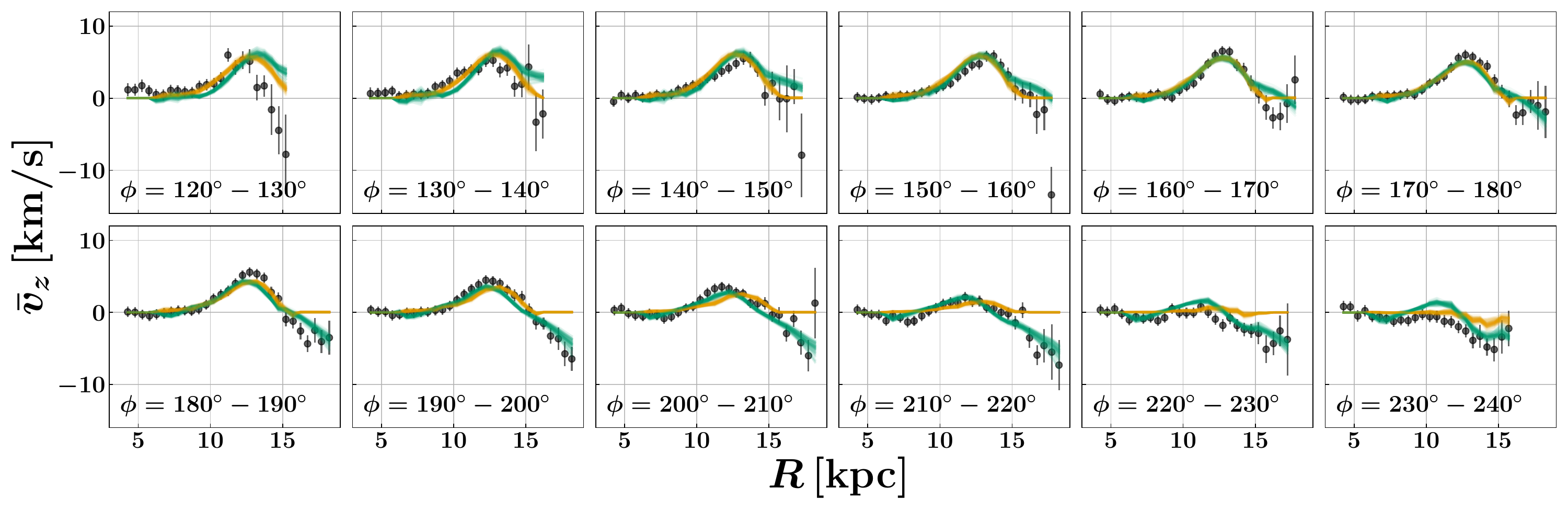}
    \caption{Fits to the $\Bar{v}_z$ data for the \emph{$v_R$ term} models. Like Fig~\ref{fig:Drimmel_fit}, each panel shows data for $10^\circ$ sectors of the Galaxy, and plots the average $v_z$ in $0.5\kpc$ bins, with associated uncertainties (statistical and systematic). Note that the uncertainties are larger here because we include uncertainties propagated from the $v_R$ uncertainties. The models with a single value of $\phi_w$ and $\omega_p$ are plotted in orange, while those of the \emph{twisted} models are in green. Both models provide reasonable fits, but note in particular that model $\Bar{v}_z$ values are all around zero at larger $R$.}
     \label{fig:Cheng_fit}%
 \end{figure*}

We can create a quantitive model for the expected velocities in idealized models using Jeans' first equation \citep{jeans_theory_1915}, which is an integral of the collisionless Boltzmann equation to give, effectively, the continuity equation. 
In cylindrical polar coordinates $(R,z,\phi)$ we can write it as 
\begin{equation} \label{eq:Jeans_1st}
    0 = \frac{\partial n}{\partial t} + \frac{\partial (n \Bar{v}_R)}{\partial R} + \frac{1}{R} \frac{\partial (n \Bar{v}_\phi)}{\partial\phi} + \frac{\partial(n \Bar{v}_z)}{\partial z},
\end{equation}
where $n$ is the number density of stars in the disc. We assume that $n$ follows a double exponential profile for stars in the disc
\begin{equation}\label{eq:double_exponential_profile}
    n(R, z') = f(R)g(z') = n_0 \exp\left(-\frac{|z'|}{z_h} - \frac{R}{R_h} \right),
\end{equation}
where the scale height $z_h$ and scale length $R_h$ are parameters that ultimately disappear at the end of the model derivation.
The term $z'$ is the distance along the $z$-axis to the warped midplane of the disc at $z_0$ (Eq.~\ref{eq:z0_simple}), and is given by
\begin{equation}
    z' = z - z_0 = z - h(R)s(\phi, R).
\end{equation}

The standard approach to warp modelling in the Milky Way has been to assume a rigidly processing warp, and that the effect of $v_R$ is negligible, yielding \citep{drimmel_galactic_2000,poggio_evidence_2020}
\begin{equation}\label{eq:standard_model}
        \Bar{v}_{z,{\rm mod}} =  \left(\frac{\Bar{v}_\phi}{R} - \omega_p\right)h(R) \frac{\partial s}{\partial \phi}.
\end{equation}

However, as we noted above, \cite{cheng_exploring_2020} showed how to include the effect of $v_R$ motion in this equation, and we can follow their derivation, with the same assumptions that $\partial\Bar{v}_\phi/\partial\phi=0$ and $\partial\Bar{v}_z/\partial z = 0$, to arrive at an expression for $\Bar{v}_z$

\begin{equation}\label{eq:cheng_model}
\begin{split}
    \Bar{v}_{z,{\rm mod}} = &\: \left(\frac{\Bar{v}_\phi}{R} - \omega_p\right)h(R) \frac{\partial s}{\partial \phi}
    \;\; + \;\;\Bar{v}_R\frac{\partial h(R)}{\partial R}s(\phi, R)\\
    & \;\; + \;\; \Bar{v}_R h(R) \frac{\partial s(\phi, R)}{\partial R}.
\end{split}
\end{equation}

This differs from the equation derived by \cite{cheng_exploring_2020} (their eq. 9), by the addition of the final term which includes $\frac{\partial s(\phi, R)}{\partial R}$. This new term is required when we introduce the twist to the line of nodes and precession rate as given in eqs.~\ref{eq:s2}~\&~\ref{eq:wpLONpoly}.

Again, we note that the parameters of the model, such as $h(R)$ and $\phi_w$, are properties of a kinematic model which, if the model is correct, correspond to the physical height of the warp and its line of nodes. The model is inevitably incomplete, and there are clear degeneracies between $h$ and $\omega_p$ implicit in eqn.~\ref{eq:standard_model}. We draw a distinction between these model parameters and these properties measured in other studies by referring to the physical warp height and line of nodes as $h_{phys}$ and $\phi_{w,phys}$, respectively.

\subsection{Models summary}

The modelling approach described above leaves us with four different models that we fit to the data. The names we will give them through this study are:

\begin{enumerate}
    \item \emph{Classical} –- Assume that we should neglect the $v_R$ terms, and fix $\omega_p$ and $\phi_w$ as a single values (eqs.~\ref{eq:s}~\&~\ref{eq:standard_model}).
    \item \emph{Classical twisted} -- Assume that we should neglect the $v_R$ terms, but allow $\omega_p$ and $\phi_w$ to vary linearly with radius (eqs.~\ref{eq:s2}~\&~\ref{eq:standard_model}).
    \item \emph{$v_R$ term} -- Include the $v_R$ terms, and fix $\omega_p$ and $\phi_w$ as a single values (eqs.~\ref{eq:s}~\&~\ref{eq:cheng_model}).
    \item \emph{$v_R$ term twisted} -- Include the $v_R$ terms, and allow $\omega_p$ and $\phi_w$ to vary linearly with radius (eqs.~\ref{eq:s2}~\&~\ref{eq:cheng_model}).
\end{enumerate}

For all of these models we treat $\Bar{v}_z$ as the quantity of interest, for which we have measured values and their statistical uncertainties. The statistical uncertainties on $\Bar{v}_\phi$ and $\Bar{v}_R$ enter as, effectively, uncertainties in the model $\Bar{v}_{z,{\rm mod}}$ velocity,
\begin{equation}\label{eq:standard_model_uncertainty}
    \sigma_{\Bar{v}_{z,{\rm mod}}} =  \sigma_{\Bar{v}_\phi}\frac{h(R)}{R}  \frac{\partial s}{\partial \phi}
\end{equation}
and
\begin{equation}\label{eq:cheng_model_uncertainty}
        \sigma_{\Bar{v}_{z,{\rm mod}}}  = \sigma_{\Bar{v}_\phi} \frac{h(R)}{R}  \frac{\partial s}{\partial \phi}
     + \;\;\sigma_{\Bar{v}_R}\left(\frac{\partial h(R)}{\partial R}s(\phi, R) + h(R) \frac{\partial s(\phi, R)}{\partial R}\right),
\end{equation}
for the \emph{Classical} and \emph{$v_R$ term} models, respectively. We add these in quadrature to the measurement uncertainty $\Bar{v}_z$, along with a systematic uncertainty $ \sigma_{\Bar{v}_{z,{\rm sys}}}=0.7\kms$ to give us 
\begin{equation} \label{eq:total_uncertainty}
    \sigma_{\Bar{v}_{z,\text{tot}}} = \sqrt{\sigma_{\Bar{v}_{z,\text{stat}}}^2 + \sigma_{\Bar{v}_{z,\text{mod}}}^2 + \sigma_{\Bar{v}_{z,\text{sys}}}^2}.
\end{equation}
The systematic uncertainty allows for the imperfections in both the model and the data, and prevents the fit being dominated by the few bins near the Sun that contain over a million stars (see Fig.~\ref{fig:maps}) and therefore have tiny statistical uncertainties. The value of $0.7$ km s$^{-1}$ was chosen such that simple models are, on average, around 1$\sigma$ from the measurement in each bin. Changing this value by a factor of 2 in either direction does not change our results substantially.

To find the best fitting parameters of the model, and their associated uncertainties, we use a Markov Chain Monte Carlo approach using the \cite{goodman_ensample} sampling scheme as applied in the Python package \texttt{emcee} \citep{foreman-mackey_emcee_2013}. The log-probability that we sample is then simply
\begin{equation}\label{eq:log_prob}
    \ln p = -\frac{1}{2} \sum \frac{(\Bar{v}_{z,\text{model}} - \Bar{v}_{z,\text{data}})^2}{\sigma_{\Bar{v}_{z,\text{tot}}}^2} + \const
\end{equation}
which is equivalent to taking a flat prior on all of our parameters. We run 50 walkers for 20\,000 steps, discarding the first 20 percent as burn-in. 



\section{Fit to the data} \label{sec:fit_to_data}

\begin{figure}
    \centering
    \includegraphics[width=\hsize]{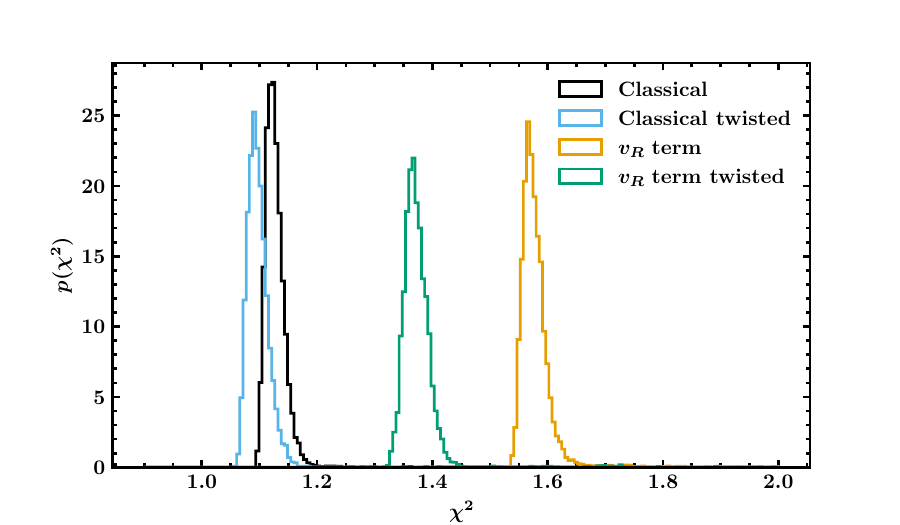}
       \caption{Quantitative comparison of the four models. We compute a quantity analogous to the  chi-squared statistic per degree of freedom, $\chi^2_\nu$, for each model, and plot the distribution of these values for the final 50\,000 steps of the MCMC chain. We note that the \emph{Classical} models perform significantly better than the \emph{$v_R$ term}.
               }
          \label{fig:chi-square}
    \end{figure}

The models we use in this study have 10 parameters that describe the greatest height of the warp as a function of $R$ ($h_7, ..., h_{16}$) and either 2 or 4 which describe the position of the line of nodes and the pattern speed of the warp, with only the two \emph{twisted} models having 4 parameters (eqs.~\ref{eq:wpLONpoly}). 

Appendix~\ref{app:parametervalues} gives the best fitting values of the parameters and their uncertainties in numerical form, while we present them graphically in this main body of the paper. In Figs.~\ref{fig:Drimmel_fit}~\&~\ref{fig:Cheng_fit} we show the fits to the data. In these figures, as in all figures showing model parameters, we represent the uncertainty by randomly drawing 100 samples of the MCMC chain and plotting them as semi-transparent lines. The uncertainties quoted in Appendix~\ref{app:parametervalues} are taken from the $15.9$ and $84.1$ percentiles of the MCMC chain after burn-in. Experiments in which we limit our sample to bins with 1000 or more stars give results that are very similar to those shown here, except with larger uncertainties, and these results are show as an additional table in the appendix.

All four kinematic models capture at least some of the features of the data, 
with the characteristic rise and fall of 
$\Bar{v}_z$ as seen by 
\cite{cheng_exploring_2020} found from $\phi=120^\circ$ to ${\sim}210^\circ$. 
For the \emph{Classical} models the data for $\phi\gtrsim210$ values is not perfectly fit by the model. The model flattens towards zero in the outer radii, while the data is almost all below zero. This is somewhat improved for the \emph{twisted} model, but not for the bins with the largest values of $\phi$. For the \emph{$v_R$ term} models the model $\Bar{v}_z$  flatten out to zero in the outer radii in a way that the data does not. The \emph{twisted} versions of these models provide the best fit of all our models at $\phi>210$, but this comes at the cost of the worst fit of all the models at low $\phi$. 

In Fig.~\ref{fig:chi-square} we show a quantitative comparison of the quality of fit for the four models. We compute a quantity analogous to the  chi-squared statistic per degree of freedom, $\chi^2_\nu$, for each model
\begin{equation}\label{eq:chisq}
    \chi^2_\nu = \frac{1}{N_{\rm bin} - N_{\rm param}} \sum \frac{(\Bar{v}_{z,\text{model}} - \Bar{v}_{z,\text{data}})^2}{\sigma_{\Bar{v}_{z,\text{tot}}}^2} 
\end{equation}
where $N_{\rm bin} = 327$ is the number of bins for our data and the number of parameters $N_{\rm param} = 12$ or $14$ for the different models. We emphasise that setting the systematic uncertainty that we apply to be $0.7\kms$ was guided by the premise that this would lead to $\chi^2_\nu\approx1$ for simple models, so the absolute value here is far less important than the relative values for the different models. The \emph{Classical} models perform significantly better than the \emph{$v_R$ term} models. We also see that the \emph{twisted} models perform better than the single value models, but that the improvement is not as great as the difference between the \emph{$v_R$ term} and \emph{Classical} models. These relative results are not changed if we change the systematic uncertainty by a factor of 2 in either direction. 

In summary, while all four model provide fits to the kinematic data that seem reasonable to the eye, a quantitative comparison shows that the \emph{Classical} models perform significantly better than the \emph{$v_R$ term} ones. 

\section{Properties of the models} \label{sec:properties}

We now look at the properties of the models, bearing in mind that the \emph{Classical} models provide the better fit to the data. 

\subsection{Warp amplitude} \label{sec:warp_amplitude}
  
\begin{figure}
    \centering
    \includegraphics[width=\hsize]{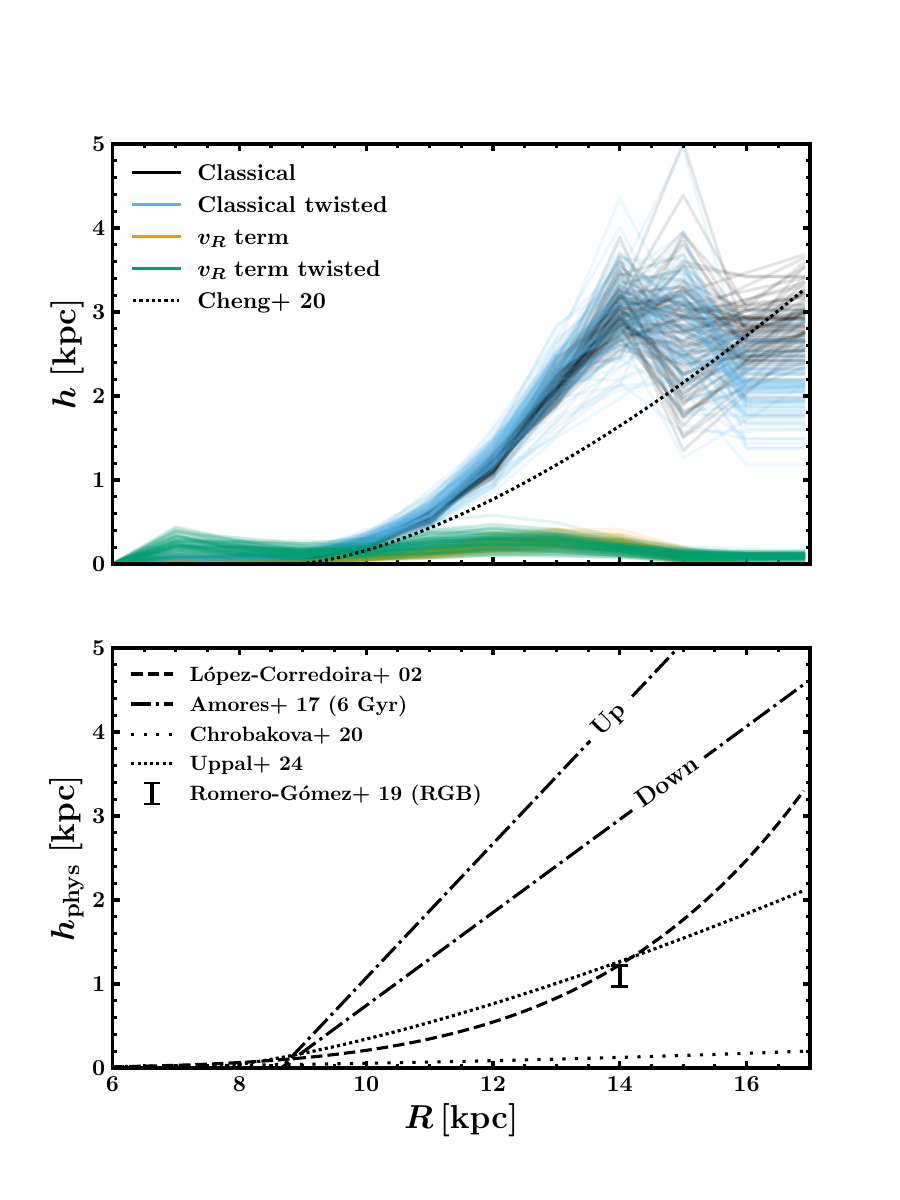}
       \caption{Maximum warp amplitude as a function of radius for our four models (top panel), and for a selection of recent estimates for the older population -- rather than in Cepheids or other young populations -- in the literature (lower panel). Two of these studies looked at the lopsidedness of the warp and in these cases we give both the warp upwards and downwards, indicated by the two lines for \citeauthor{amores_evolution_2017} (\citeyear{amores_evolution_2017}, with the upwards warp being the greater) and the uncertainty bar for the Red Giant Branch (RGB) results of \citeauthor{romero-gomez_gaia_2019} (\citeyear{romero-gomez_gaia_2019}, with the downwards warp being the greater value). 
               }
          \label{fig:warp_amplitude}
    \end{figure}

In Fig.~\ref{fig:warp_amplitude} we show the maximum amplitudes of the warp in our models, and how it compares with other measurements of the warp amplitude found in the literature. We divide this into results that were found purely from kinematics (labelled $h$) and from photometric data and star counts (labelled $h_{phys}$). The two \emph{Classical} models reach ${\sim}3\kpc$ at $R\approx14\kpc$, and then remain around the same height (with significant uncertainty) further out. The two \emph{$v_R$ term} models both stay at much smaller heights, reaching only ${\lesssim} 0.3\kpc$ by $R\approx13\kpc$ and then noticeably decreasing further out, down to almost zero in the case of the untwisted model. 

The literature values we show come from studies which looked at the bulk of the population or specified older populations \citep{lopez_corredoira_2mass_2002,amores_evolution_2017,cheng_exploring_2020,chrobakova_structure_2020,romero-gomez_gaia_2019,uppal_clump_2024}, as opposed to those which looked at the gas or young populations, such as Cepheid variables, that tend to be less warped than all of these except the \cite{chrobakova_structure_2020} estimate \citep[$h$ typically around $0.5{-}0.8\kpc$ at $R=14\kpc$ for Cepheids, e.g.,][]{skowron_mapping_2019,dehnen_a_twisted_2023,cabrera_cepheids_2024}. We pick out specifically the results using red-giant branch (RGB) stars from \cite{romero-gomez_gaia_2019}. We chose results for $6\Gyr$ old stars from \cite{amores_evolution_2017}, corresponding to their second-oldest age bin. Their oldest age bin ($8.3\Gyr$) has significantly larger uncertainties and even steeper warps. Of the results shown, only ours and that from the \cite{cheng_exploring_2020} study comes purely from kinematic data, while the other studies used photometric data and star counts to trace the warp shape.

 The \emph{Classical} models have a warp amplitude that lies somewhere in the middle of these models, a stronger warp at ${\sim}14\kpc$ than in all but the \cite{amores_evolution_2017} case, but flattening out to be fairly similar to both the \cite{lopez_corredoira_2mass_2002} and \cite{cheng_exploring_2020}. This flattening is not something one would expect for a typical warp, and does come with significant uncertainties. The \emph{$v_R$ term} models have a much smaller amplitude than all but the \cite{chrobakova_structure_2020} warp, 
 and do not look like a warp in the sense that they do not tend towards greater heights at greater radii. 
 
 Our \emph{Classical} models are close to the results from \cite{cheng_exploring_2020}, which in one sense is unsurprising, as we both fit the kinematic data alone. However, one might have expected the results from our \emph{$v_R$ term} models to be more similar to theirs, since they also used these models in their analysis. The reason for this difference is probably that we analysed the data in several azimuthal bins at each radius, while \citeauthor{cheng_exploring_2020} combined data at all azimuths into a single radial bin, thus reducing the effect of the $v_R$ term on their model.

Our robustness tests in which we cut the data at different maximum values of $R$ produces very similar height profiles to those found here for all models, and we show these in the appendix. Our robustness tests in which we look at $2\kpc$ wide rings in $R$ has slightly low values of $h$ at $R=16$ in the \emph{Classical} models, with large uncertainties. The results from the \emph{$v_R$ term} models looking at $2\kpc$ wide rings have greater $h$ than those we see here, but this reflects a weakness of this piecewise-linear model applied to this limited sample, rather than a worrying feature of the data analysis. We discuss this, and show results from all of these robustness tests in Appendix~\ref{app:parametervalues}.

In summary, the \emph{Classical} models, in addition to providing a better fit to the data, also provide a warp amplitude that is more in line with the literature. The \emph{$v_R$ term} models, as well as being a worse fit for the data, do not resemble warps as we would expect them. We draw the conclusion that the \emph{Classical} models are much more likely to be a helpful representation of the warp in the Milky Way.

\begin{figure}
   \centering
   \includegraphics[width=\linewidth]{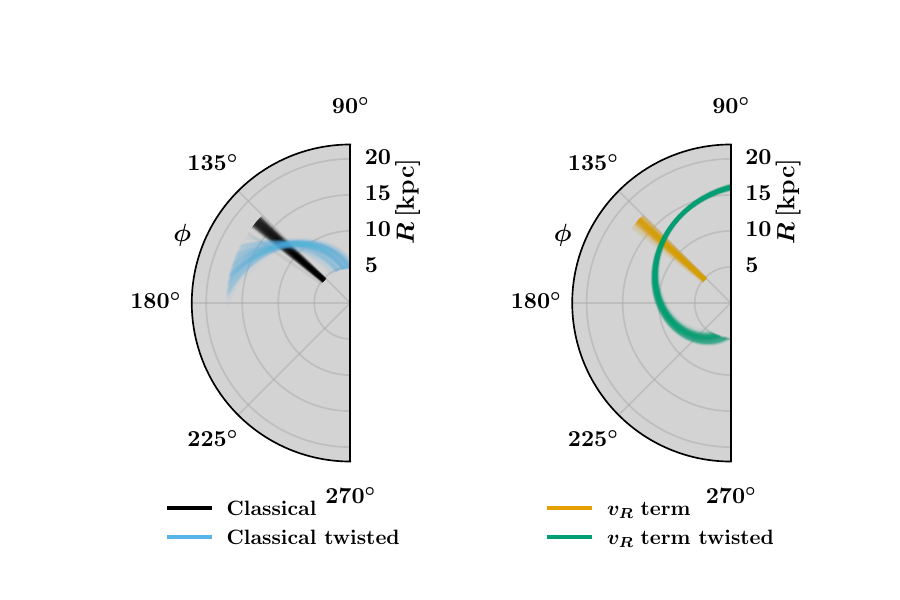}
   \caption{Position of the line of nodes for our \emph{Classical} (left) and \emph{$v_R$ term} models. The \emph{Classical twisted} model has a line of nodes that forms a trailing spiral, while the \emph{$v_R$ term twisted} model's line of nodes is a leading spiral that is much more wound up.}
    \label{fig:LON}%
\end{figure}

\begin{figure}
    \centering
    \includegraphics[width=\linewidth]{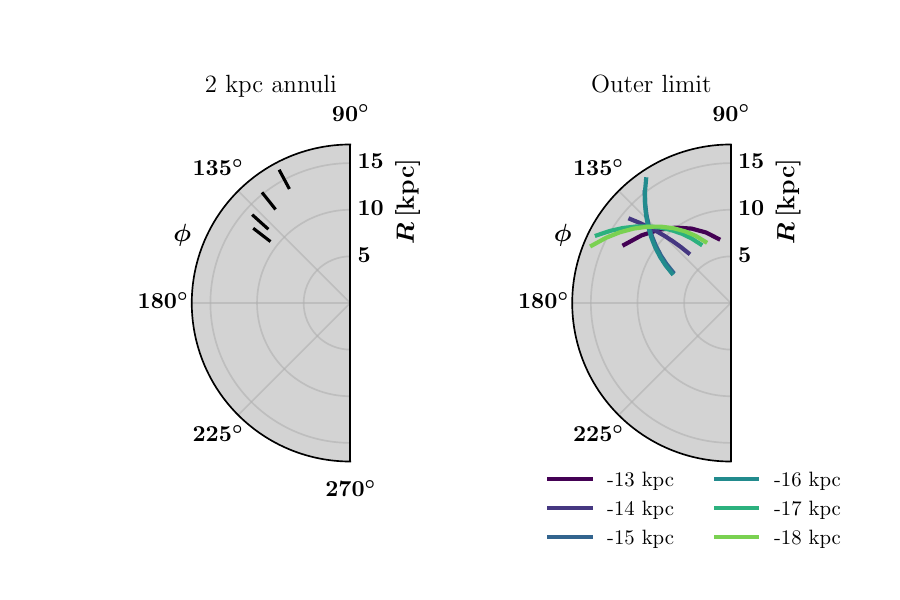}
    \caption{Position of the line of nodes for two of our robustness tests. The left panel shows the results of our \emph{Classical} model when we cut the data into $2\kpc$ wide rings from $11{-}13\kpc$ to $14{-}16\kpc$. The right panel shows our \emph{Classical twisted} model when we cut the data at different maximum radii from $13$ to $18\kpc$. To avoid clutter in the figures we only show the line of nodes for the best fitting model for each sample. It is worth noting that in the right panel the models for data with $R<15$ or $16\kpc$ are so similar that they sit on almost perfectly top of each other. The different trends seen for the different samples make it hard to draw concrete conclusions about the twist of the line of nodes.}
     \label{fig:robustLON}%
 \end{figure}

\begin{figure}
   \centering
   \includegraphics[width=\linewidth]{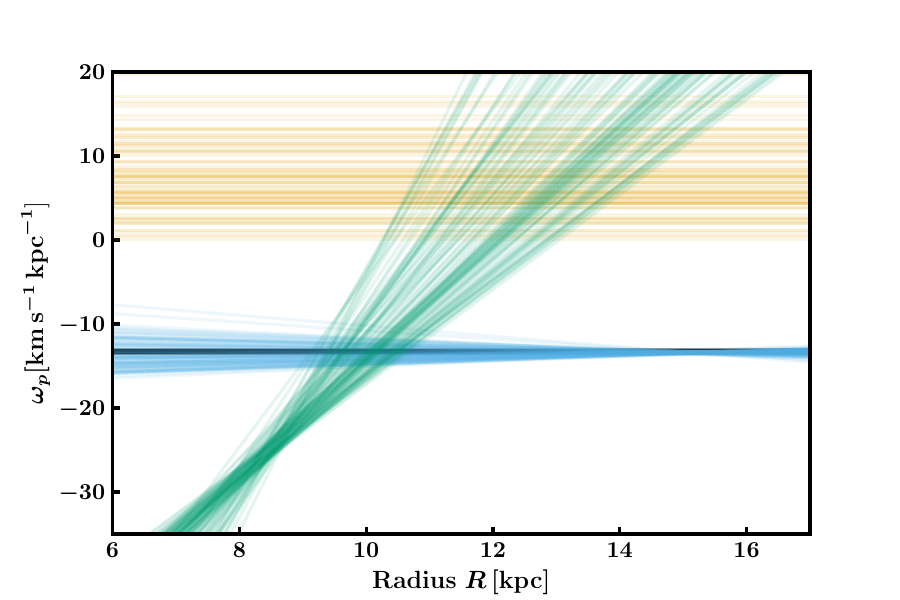}
   \caption{Precession of the line of nodes for our 4 models. Note that the disc is spinning in the negative sense, so the two \emph{Classical} models are both precessing in the prograde sense, with very similar precession rates. The \emph{Classical twisted} model has negligible change in precession rate with radius. The two \emph{$v_R$ term} models have very different precession rates with the \emph{$v_R$ term twisted} model having a dramatic change in precession rate with radius.}
    \label{fig:wp}%
\end{figure}

\subsection{Line of nodes} \label{sec:line_of_nodes}
The positions of the line of nodes for each of our four models are shown in Fig.~\ref{fig:LON}. We plot it from $R=6\kpc$ outwards, as our model warp has zero amplitude inside this radius. In both the cases where the line of nodes is held at a fixed value, it lies around $\phi=140^\circ$. This reflects the clear signature in the data (Fig.~\ref{fig:maps}) that towards $\phi=230^\circ$ the vertical velocity is close to zero. The \emph{Classical twisted} model has a line of nodes that forms a trailing spiral, while the \emph{$v_R$ term twisted} model's line of nodes is a leading spiral that is much more tightly wound. In both cases these cross $135^\circ$ at $R\approx 12\kpc$, which is the radius where the velocity signal of the warp appears to be strongest.

This line of nodes is significantly different from that found in the literature using the current positions of stars. Most early 21st-century studies of the Milky Way's warp place $\phi_{w,phys}$ very close to $180^\circ$ \citep[e.g.][]{drimmel_galactic_2000,lopez_corredoira_2mass_2002,levine_vertical_2006}, but some more recent work has placed it closer to $165^\circ$ in both the Cepheid population \citep[e.g.][]{skowron_mapping_2019} and the older population \citep[e.g.][]{momany_outer_2006,amores_evolution_2017}. 

Both \cite{chen_intuitive_2019} and \cite{dehnen_a_twisted_2023} find a line of nodes for samples of Cepheids that is close to $180^\circ$ for $R = 12\kpc$, but then decreases almost linearly in $\phi$ (a leading spiral) to $140^\circ$ around $R=16\kpc$. This is measured using the positions of the Cepheids by both studies, while \citeauthor{dehnen_a_twisted_2023} find the same result again when they also use the velocities to measure the warp's properties from the angular momenta of the stars around the Galactic centre\footnote{This is neither a measurement using purely physical positions nor purely kinematics}. Inside $12\kpc$, the warp measured using only the positions of the Cepheids is instead a trailing spiral in both studies. The warp measured by \cite{dehnen_a_twisted_2023} using angular momentum is generally too small inside $12\kpc$ to draw any conclusions about its line of nodes there. 

Perhaps most relevantly for understanding our results, using Gaia DR2 \cite{romero-gomez_gaia_2019} find that, for red giant stars, their estimates of the position of the line of nodes using the positions ($\phi_{w,phys}{\sim}180^\circ-200^\circ$) and kinematics ($\phi_w{\sim}160^\circ-170^\circ$) of the stars are rather different from one another. They attributed this to the warp being lopsided, with the line of nodes being offset from the maximum vertical velocity.


The robustness tests we have conducted suggest that the trailing spiral of the line of nodes seen in the \emph{Classical twisted} model is not a secure result from our data. In Fig~\ref{fig:robustLON} we show two examples of results from these tests which illustrate this (more complete figures can be found in Appendix~\ref{app:parametervalues}). When we divide the data into $2\kpc$ annuli and fit a \emph{Classical} model to each, we find that the different lines of nodes appear to be forming a leading spiral. When we vary the outermost radius of the data we fit, we find that the twist of the line of nodes swings between trailing for the $R<13$ and $14\kpc$ samples, to leading for the $R<15$ and $16\kpc$ samples, then back to trailing for the $R<17$ and $18\kpc$ samples. We therefore can not make a strong statement about the twist of the line of nodes. 

We can however say with some confidence that the kinematically estimated line of nodes based on this population is clearly leading that found in the positions of similar samples (or from Cepheids) and that this result is robust against selection of different subsets of our data.

\subsection{Precession rate} \label{sec:precession_rate}
Figure \ref{fig:wp} shows the precession rate of the line of nodes for each of our four models. The \emph{Classical} models have a precession rate of $\omega_p \approx -13.5\kms\kpc^{-1}$ (prograde with the disc rotation). This would be equivalent of a full rotation around the galaxy in around half a Gyr. The \emph{Classical twisted} model has negligible change in precession rate with radius. 

This relatively fast prograde precession of the warp is consistent with the results from \cite{poggio_evidence_2020} and \cite{cheng_exploring_2020}. It is consistent with results for Cepheids from \cite{dehnen_a_twisted_2023} at $R\approx12\kpc$, but they found that the precession speed decreased to ${\sim}6\kms\kpc^{-1}$ for $R\gtrsim13\kpc$. Our \emph{Classical twisted} model allows for this possibility, but does not find it.

The \emph{$v_R$ term} model has a retrograde precession of around $8\kms\kpc^{-1}$, while the \emph{$v_R$ term twisted} model has a dramatic change in precession rate with radius of ${\sim}3\kms\kpc^{-2}$, changing sign from prograde at lower radii to retrograde for $R\gtrsim12\kpc$. This seems rather unlikely, and we consider it be a consequence of the fitting algorithm hunting down a mathematically acceptable solution for a model where a physically meaningful one was not available.

Our robustness tests reinforce these results, with the tests in which we set a maximum $R$ value all being consistent with these values. The tests with $2\kpc$ wide rings in $R$ are the same for the \emph{Classical} models, but the \emph{$v_R$ models} become more consistent with the \emph{Classical} models. Like the equivalent for $h(R)$ in these models, this reflects a weakness of this piecewise-linear model applied to this limited sample.

\subsection{Results summary} \label{sec:results_summary}
The \emph{Classical} models provide a better fit to the data, with an amplitude and shape consistent with typical warps, and a precession speed consistent with other measures in the literature. Their line of nodes does not lie where one would expect given results in the literature based on the positions of stars. The \emph{$v_R$ term} models provide a worse fit to the data, with an amplitude and shape that are not consistent with typical warps and, in the \emph{twisted} case, a precession speed that varies dramatically with radius.

\section{Discussion} \label{sec:discussion}

\subsection{The $v_R$ term} \label{sec:vr_term}
\cite{cheng_exploring_2020} introduced the idea that one could include the velocity of stars radially in the disc ($v_R$) in the kind of Jeans modelling of the warp that we have used here. While they had a smaller dataset, and combined stars at all azimuths ($\phi$) into a single bin at each radius, we have exploited the larger dataset from \Gaia\ DR3 to look at the vertical velocity of the warp as a function of $\phi$ as well as $R$. 

It is important to recognize that the magnitude of the $v_R$ terms' effect on the model vertical velocity (eq.~\ref{eq:cheng_model}) is $90^\circ$ out of phase with the effect of the other term (the only term in the \emph{Classical} models, eq.~\ref{eq:standard_model}). The \emph{Classical} models have greatest $\Bar{v}_{z,{\rm mod}}$ at the line of nodes because their value depends on the derivative of warp height with $\phi$, which is greatest there. The $v_R$ terms, in contrast, have their maximum effect at the $\phi$ value where the warp height is maximum, because their value depends on the derivative of warp height with $R$. This means that the influence of the $v_R$ terms is zero at the line of nodes.

Our results with the \emph{Classical} models are very similar to those of \cite{cheng_exploring_2020}, while those with the \emph{$v_R$ term} models are very different. The most likely explanation for this is that when the \cite{cheng_exploring_2020} study combined all azimuths into a single bin at each radius, they treated the stars as all being at the median azimuth of the bin. This will be close to $180^\circ$ which they assumed was the location of the line of nodes. This means that the $v_R$ terms will have had a small effect in their fits. In contrast, because we have fit the data in bins of $\phi$ at each radius, the $v_R$ terms have a much greater effect on our fits across the disc. 

We should be a little cautious, because as \cite{gaia_drimmel_2023} point out, $v_R$ estimates are prone to systematic errors across the disc. We can expect that our use of the \cite{bailer-jones_estimating_2021} Bayesian distances has reduced these errors, and since our robustness tests with much more local samples come to similar results, we are confident that including these $v_R$ terms makes the fit to the data worse, even when we have allowed the properties to the warp to vary with radius in a way that, especially for the precession of the warp, seems physically implausible. It seems therefore that including these terms makes our model of the warp significantly worse.

The radial motions of stars are not the only things omitted from our \emph{Classical} models that might be important for analysing large scale trends due to the warp. We ignore any possible variation of warp amplitude with time, and our variation with $\phi$ is a simple sinusoid. \cite{cabrera_cepheids_2024} found that for a sample of Cepheids the lopsidedness of the warp (parameterized as an $m=2$ mode) was an important factor, while the change in amplitude was not. We defer investigation of these for the bulk population to future work.

As \cite{box_models_wrong_76} noted: all models are wrong, but some are useful. Our warp models cannot capture the full complexity of the warp, but they can be useful in understanding the warp's properties. The assumption underlying our models is that the warp is fixed, except for its precession. We therefore require that a star moving radially `knows about' the height of the warp at the radii it is moving towards. The radial motions only exist because the warp is not fixed, and the outer parts of the disc have a complicated structure including spiral arms. Introducing these radial motions into our models is therefore a way of trying to capture the complexity of the warp. However, it seems that this is not a successful way of doing so.


\subsection{The inconsistent line of nodes} \label{sec:lon} 
The line of nodes in our \emph{Classical} models lies near $\phi = 140^\circ$ at $R=12\kpc$. As discussed in Sect.~\ref{sec:properties} this is not the same as the line of nodes found by numerous studies of the warp in the Milky Way, whether of position of the bulk population or young populations such as Cepheids, which tend to be closer to $165^\circ-180^\circ$. 

From this we conclude that the observed structure and kinematics of the bulk population in the outer disc are inconsistent with the picture of a fixed, precessing, warped disc. This stands in contrast to the younger Cepheid population for which \cite{dehnen_a_twisted_2023} found warps defined by the positions or angular momenta of the stars were much more consistent -- though not perfectly consistent, as they discuss. We note also that \cite{cabrera_cepheids_2024} found that the line of maximum $z$ velocity for the Cepheid population, which for our model would be the line of nodes, trailed the line of nodes found from the positions of stars by ${\sim}20^\circ$, the opposite of the trend we see in the older population. 

Our approach within this paper has been to look only at the kinematics of the sample. This is a pragmatic choice, since it avoids dealing with the dust extinction within the Milky Way, but it has great scientific merit to look at the two probes (density distribution and kinematics) separately, and see if they are consistent. Doing so allows us to check our assumptions, and we can see that they do not hold up.

We are not the first to note that the line of nodes determined from kinematics is offset from the Galactic anticentre, with \cite{poggio_galactic_2018} noting that the maximum in $\Bar{v}_z$ was at $\phi<180^\circ$ and \cite{romero-gomez_gaia_2019} placing it in the range $160^\circ{-}170^\circ$. Our study places the line of nodes even further from the anti-centre than these studies. This is appears to be because we fit a model to the kinematics alone, and this increases the significance that $\Bar{v_z} \approx 0$ at $\phi\approx220^\circ$ even at $R{\sim}12\kpc$ where we see the strongest positive $\Bar{v_z}$ elsewhere in the disc. This is $90^\circ$ away from the line of nodes in our warp model.

\subsection{The disturbed Milky Way} \label{sec:disturbed}

The Milky Way's disc is not a particularly calm and settled system. The spiral arms create large disturbances in the radial velocities of stars within the disc \citep[e.g.][]{Gaia_katz_18}. Even before \Gaia\ DR2, disturbance of the vertical velocity of stars across the disc had been and found and associated with `corrugations' or `bending' and `breathing' modes of the disc \citep{widrow_bending_2014,schonrich_warp_2018}.  One of the first discoveries with Gaia DR2 was the `phase-spiral' \citep{antoja_dynamically_2018}
which is a disturbance seen in the density in the $z{-}v_z$ plane of stars in the Solar neighbourhood. Further studies have shown that this is how far from the Solar neighbourhood this feature can be found \citep[][]{xu_outer_2020}, and that it grows particularly strong when looking specifically at stars with orbital guiding centre radii around $10\kpc$ \citep[angular momentum ${\sim}2300\kms\kpc$,][]{alinder_amplitude_2023}.

Investigation of the outer disc by \cite{gaia_collaboration_gaia_2021} and \cite{mcmillan_disturbed_2022} showed that there is a bifurcation of the velocity distribution at radii near $12\kpc$, which varies as we look around the disc. This is near the peak of the warp signal in our data. 

None of this behaviour is captured by a simple warp model of the kind we have used here. Perhaps, therefore, we should not be surprised by the fact that our kinematic model of the warp is not consistent with the positions of stars in the outer disc. None-the-less, we have quantified this inconsistency and are left with quite a stringent constraint for any model of the creation of the Milky Way's warp. A range of such models have been proposed in the recent literature, most focussing on the LMC and its effect on the dark-matter halo \citep[e.g.][]{han_tilted_2023}, the Sagittarius dwarf \citep[e.g.,][]{poggio_measuring_2021} or a combination of both \citep[e.g.,][]{laporte_influence_2018}. To match what we observe in the Milky Way, these models need to have warps that appear to be precessing fast in the prograde sense, and the line of nodes in the kinematics of the bulk population must lead the line of nodes in the positions of the bulk population by a substantial amount.

In this sense, we also see a difference between the warp seen in the Cepheid population, which is young, and the bulk population, which is older. 
The results of \cite{dehnen_a_twisted_2023} and \cite{cabrera_cepheids_2024} suggest that the line of nodes in simple kinematic models of the Cepheid population are roughly consistent with that in position, or that they trail it, while our results have a line of nodes in the kinematic model that leads one found in position.
It may be that the Cepheids, a dynamically cold population, had a different response to perturbation than the dynamically hotter bulk population. Perhaps a more likely explanation is that the Cepheids more closely represent the disturbance of the gas disc from which they formed, while the bulk population is dominated by stars that were already in the disc when it was disturbed, and therefore we see the response of the collisionless component of the disc to the disturbance.


The difference between the responses to a Sagittarius-like perturber of the gas and stellar discs in a Milky-Way-like galaxy was investigated by \cite*{tepper_corrugations_2022}. They refer to the variation in $\Bar{v}_z$ for stars and gas across the disc as a corrugation, and note that they are initially in phase, but move apart after $500{-}700\Myr$. They argue that the difference between the two disturbances may help to determine the age of the perturbation. In this context the difference between the warp seen in the Cepheids and the bulk population may be a sign that the warp is due to a relatively old perturbation of the Milky Way's disc, but we will defer more quantitive statements about this to future work.


\section{Conclusions} \label{sec:conclusions}

In this study we have used \textit{Gaia} DR3 data to study the vertical motion of stars affected by the Milky Way's warp. The exquisite quality of the data allows us to fit a kinematic model to observations spanning a significant fraction of the Milky Way's disc. We have used this to investigate the position and behaviour of the warp, under the assumption that it is a fixed, precessing, warped disc.

The kinematic signature of the warp in the bulk population is consistent with a warp that is precessing at around $13\kms\kpc^{-1}$ in a prograde direction with a line of nodes around $\phi=140^\circ$, i.e., $40^\circ$ prograde of the Sun's position. We can not come to any robust conclusions about how this line of nodes varies with radius, because different subsamples of our data lead us to differing results.


Including the expected effects of radial motions on the continuity-equation-based model of the warp makes the fit to the data worse. We conclude that this is not a helpful addition to models of the warp's kinematics. 

The kinematic signature of the warp, under the assumption that it is fixed and precessing, does not line up with the position of the warp seen in star counts. We are forced to conclude that this model is not sufficient to describe the system, and that this provides a challenge to models of the formation of the warp in the Milky Way. The difference between this behaviour and that of the Cepheid population may provide a clue as to the nature and date of the last major perturbation of the Milky Way's disc.

\begin{acknowledgements}
     The authors are grateful to the anonymous referee for suggestions which strengthened this work, to Walter Dehnen for discussing his work on the warp with us, and providing us with a pre-publication copy of his paper. PM thanks Eloisa Poggio and Joss Bland-Hawthorn for helpful advice. The authors gratefully acknowledge support from project grants from the Swedish Research Council (Vetenskapr\aa det, Reg: 2017-03721; 2021-04153). Some of the computations in this project were completed on computing equipment bought with a grant from The Royal Physiographic Society in Lund.   This work has made use of data from the European Space Agency (ESA) mission
    {\it Gaia} (\url{https://www.cosmos.esa.int/gaia}), processed by the {\it Gaia}
    Data Processing and Analysis Consortium (DPAC,
    \url{https://www.cosmos.esa.int/web/gaia/dpac/consortium}). Funding for the DPAC
    has been provided by national institutions, in particular the institutions
    participating in the {\it Gaia} Multilateral Agreement. 
    This research has made use of NASA's Astrophysics Data System. 
    This paper made use of the following software packages for Python:
    \verb|Numpy| \citep{harris2020array}, 
    \verb|AstroPy| \citep{astropy:2013,astropy:2018,astropy:2022}, 
    \verb|emcee| \citep{foreman-mackey_emcee_2013}, 
    \verb|SciPy| \citep{2020SciPy-NMeth},
    \verb|Pandas| \citep{mckinney-proc-scipy-2010}, 
    and \verb|Matplotlib| \citep{Hunter:2007}.
\end{acknowledgements}

%
   \bibliographystyle{aa} 
   \bibliography{references} 
%

\begin{appendix} 

\section{Best fitting parameter values and robustness tests} \label{app:parametervalues}

\begin{table*}
    \centering
    \begin{tabular}{ccccc}
     Parameter & \emph{Classical} & \emph{Classical twisted} & \emph{$v_R$ term} & \emph{$v_R$ term twisted} \\\hline 
     $h_{7}$ [kpc] & $0.019 \pm 0.012$ & $0.046 \pm 0.022$ & $0.0112 \pm 0.0062$ & $0.21 \pm 0.10$\\
     $h_{8}$ [kpc] & $0.029 \pm 0.017$ & $0.063 \pm 0.027$ & $0.0112 \pm 0.0069$ & $0.168 \pm 0.073$\\
     $h_{9}$ [kpc] & $0.075 \pm 0.025$ & $0.103 \pm 0.033$ & $0.0245 \pm 0.0091$ & $0.127 \pm 0.064$\\
     $h_{10}$ [kpc] & $0.231 \pm 0.035$ & $0.263 \pm 0.050$ & $0.066 \pm 0.016$ & $0.154 \pm 0.060$\\
     $h_{11}$ [kpc] & $0.557 \pm 0.050$ & $0.615 \pm 0.089$ & $0.133 \pm 0.030$ & $0.230 \pm 0.092$\\
     $h_{12}$ [kpc] & $1.171 \pm 0.074$ & $1.26 \pm 0.16$ & $0.225 \pm 0.052$ & $0.266 \pm 0.096$\\
     $h_{13}$ [kpc] & $2.11 \pm 0.13$ & $2.19 \pm 0.28$ & $0.262 \pm 0.059$ & $0.245 \pm 0.081$\\
     $h_{14}$ [kpc] & $3.02 \pm 0.29$ & $2.96 \pm 0.43$ & $0.224 \pm 0.057$ & $0.179 \pm 0.058$\\
     $h_{15}$ [kpc] & $2.78 \pm 0.66$ & $2.82 \pm 0.63$ & $0.088 \pm 0.036$ & $0.100 \pm 0.042$\\
     $h_{16}$ [kpc] & $2.70 \pm 0.31$ & $2.13 \pm 0.41$ & $0.0023 \pm 0.0024$ & $0.090 \pm 0.030$\\
     $\phi_{w,0}$ [deg] & $139.1 \pm 1.7$ & $137.5 \pm 1.8$ & $138.0 \pm 1.6$ & $148.2 \pm 2.0$\\
     $\phi_{w,1}$ [deg kpc$^{-1}$] &  & $5.2 \pm 1.4$ &  & $-14.51 \pm 0.88$\\
     $\omega_{p,0}$ [$\mathrm{km\,s^{-1}\,kpc^{-1}}$] & $-13.26 \pm 0.11$ & $-13.45 \pm 0.52$ & $9.6 \pm 7.2$ & $6 \pm 12$\\
     $\omega_{p,1}$ [$\mathrm{km\,s^{-1}\,kpc^{-2}}$] &  & $0.02 \pm 0.16$ &  & $8.5 \pm 2.9$\\
    \end{tabular}
    \caption{Parameters of the best fitting models for our main sample with associated uncertainties. These results are represented graphically in the figures in the main text. Note that, for the twisted models, $\phi_{w,0}$ and $\omega_{p,0}$ are the values at $R=12\kpc$ for the line of nodes and precession rate respectively.}
    \label{tab:best-fit-parameters}
 \end{table*}
 \begin{table*}
     \centering
     \begin{tabular}{ccccc}
     Parameter & \emph{Classical} & \emph{Classical twisted} & \emph{$v_R$ term} & \emph{$v_R$ term twisted} \\\hline 
     $h_{7}$ [kpc] & $0.020 \pm 0.012$ & $0.047 \pm 0.027$ & $0.0070 \pm 0.0042$ & $0.30 \pm 0.11$\\
     $h_{8}$ [kpc] & $0.030 \pm 0.017$ & $0.064 \pm 0.035$ & $0.0073 \pm 0.0046$ & $0.087 \pm 0.056$\\
     $h_{9}$ [kpc] & $0.078 \pm 0.026$ & $0.121 \pm 0.045$ & $0.0156 \pm 0.0066$ & $0.093 \pm 0.047$\\
     $h_{10}$ [kpc] & $0.235 \pm 0.037$ & $0.327 \pm 0.078$ & $0.040 \pm 0.012$ & $0.161 \pm 0.056$\\
     $h_{11}$ [kpc] & $0.581 \pm 0.054$ & $0.80 \pm 0.16$ & $0.076 \pm 0.023$ & $0.215 \pm 0.071$\\
     $h_{12}$ [kpc] & $1.207 \pm 0.096$ & $1.59 \pm 0.29$ & $0.126 \pm 0.039$ & $0.254 \pm 0.076$\\
     $h_{13}$ [kpc] & $2.27 \pm 0.20$ & $2.84 \pm 0.51$ & $0.146 \pm 0.045$ & $0.225 \pm 0.063$\\
     $h_{14}$ [kpc] & $3.39 \pm 0.49$ & $3.79 \pm 0.73$ & $0.126 \pm 0.041$ & $0.165 \pm 0.048$\\
     $h_{15}$ [kpc] & $3.3 \pm 1.2$ & $2.5 \pm 1.2$ & $0.045 \pm 0.028$ & $0.055 \pm 0.031$\\
     $h_{16}$ [kpc] & $2.6 \pm 1.7$ & $2.4 \pm 1.7$ & $0.021 \pm 0.019$ & $0.021 \pm 0.020$\\
     $\phi_{w,0}$ [deg] & $138.6 \pm 1.8$ & $138.7 \pm 2.0$ & $138.6 \pm 1.6$ & $139.8 \pm 1.8$\\
     $\phi_{w,1}$ [deg kpc$^{-1}$] &  & $4.1 \pm 2.4$ &  & $-1.6 \pm 2.1$\\
     $\omega_{p,0}$ [$\mathrm{km\,s^{-1}\,kpc^{-1}}$] & $-13.46 \pm 0.25$ & $-14.42 \pm 0.65$ & $32 \pm 15$ & $7.0 \pm 9.4$\\
     $\omega_{p,1}$ [$\mathrm{km\,s^{-1}\,kpc^{-2}}$] &  & $0.35 \pm 0.24$ &  & $7.4 \pm 1.8$\\
 \end{tabular}
 \caption{For tests with a requirement that only spatial bins with 1000 stars or more are included in the fit, this table shows the parameters of the best fitting models with uncertainties. The results are consistent with our main model, but with significantly larger uncertainties in the outer disc.}
 \label{tab:n1000-best-fit-parameters}
 \end{table*}
 
We present the best fitting parameters from all of our models in Table~\ref{tab:best-fit-parameters}. These correspond to the values that we illustrate graphically in Figs.~\ref{fig:warp_amplitude}, \ref{fig:LON}, and \ref{fig:wp}. 

The results of our robustness test with a minimum of 1000 stars per spatial bin are in Tab.~\ref{tab:n1000-best-fit-parameters}, and it is clear that the results are consistent with the full sample, but with much larger uncertainties in the outer bins and in $\omega_p$ for the \emph{$v_R$ term} models. An increase in uncertainty is, of course, to be expected since there is less data to constrain the model.

\begin{figure*}
    \centering
    \includegraphics[width=0.32\linewidth]{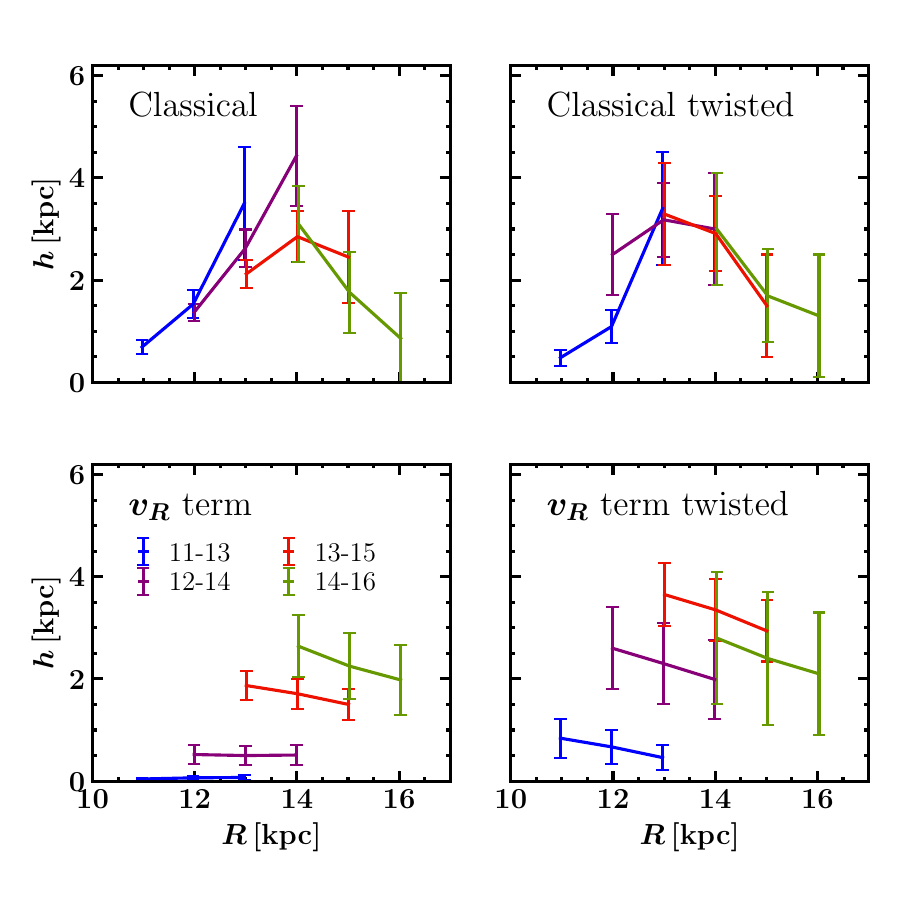}
    \includegraphics[width=0.32\linewidth]{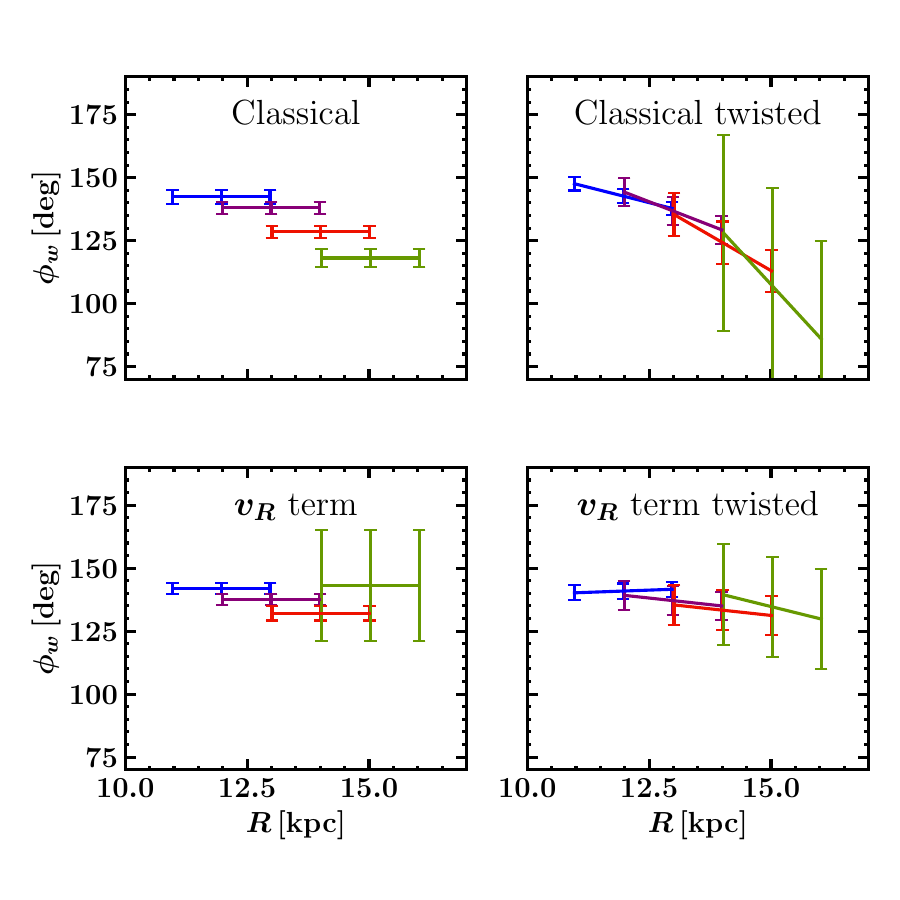}
    \includegraphics[width=0.32\linewidth]{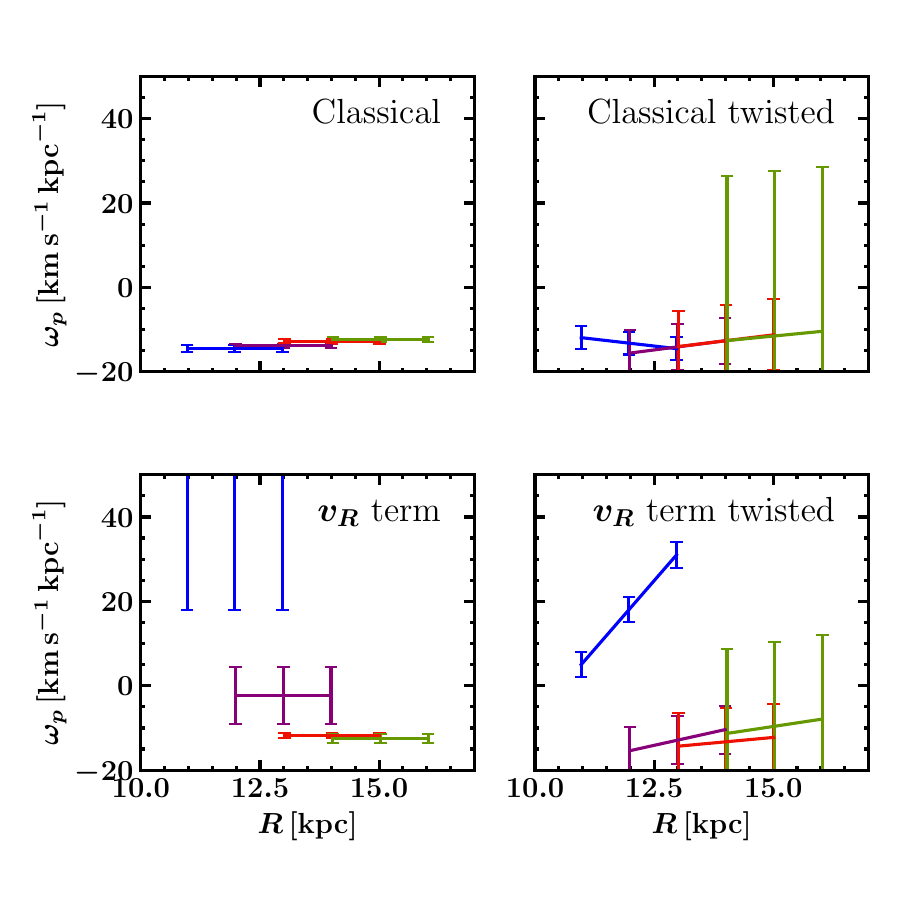}
    \caption{Maximum warp heights ($h$ -- left group of panels), line of nodes ($\phi_w$ -- centre group  of panels) and warp pattern speed ($\omega_p$, right group of panels) as a function of radius for our four models, where the data has been binned into overlapping $2\kpc$ rings in $R$, as indicated. Note that the line of nodes appears to be twisting in a prograde direction in all cases.}
     \label{fig:annuli}%
 \end{figure*}

\begin{figure*}
    \centering
    \includegraphics[width=0.32\linewidth]{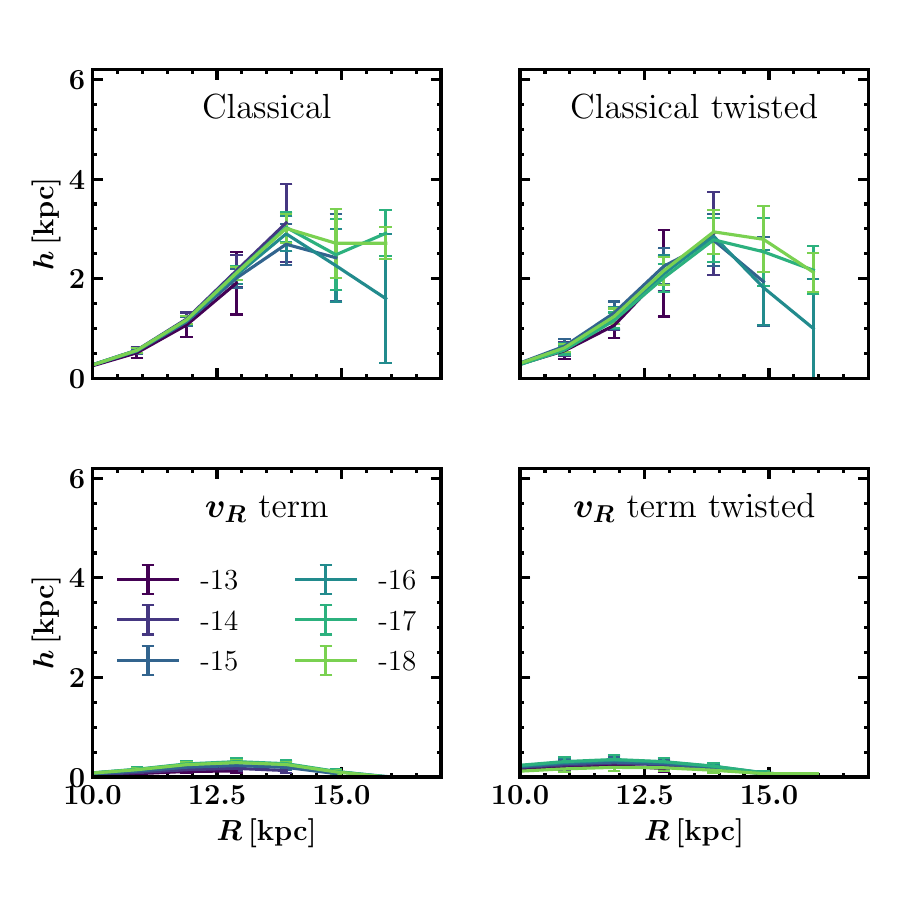}
    \includegraphics[width=0.32\linewidth]{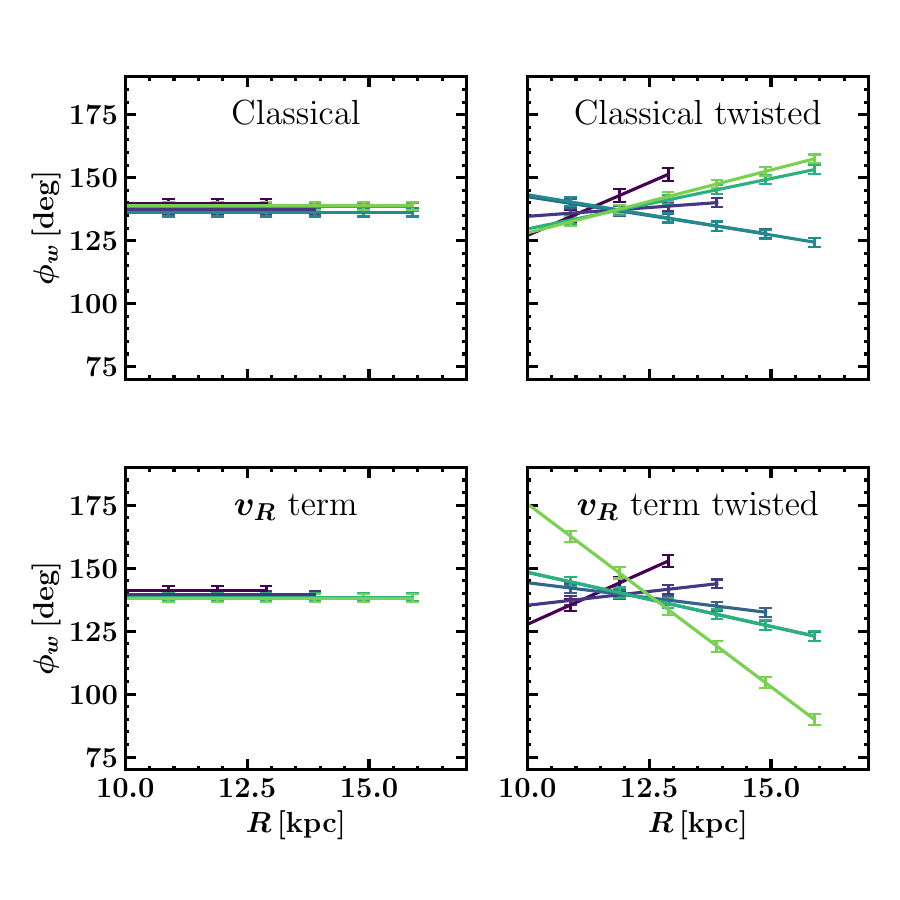}
    \includegraphics[width=0.32\linewidth]{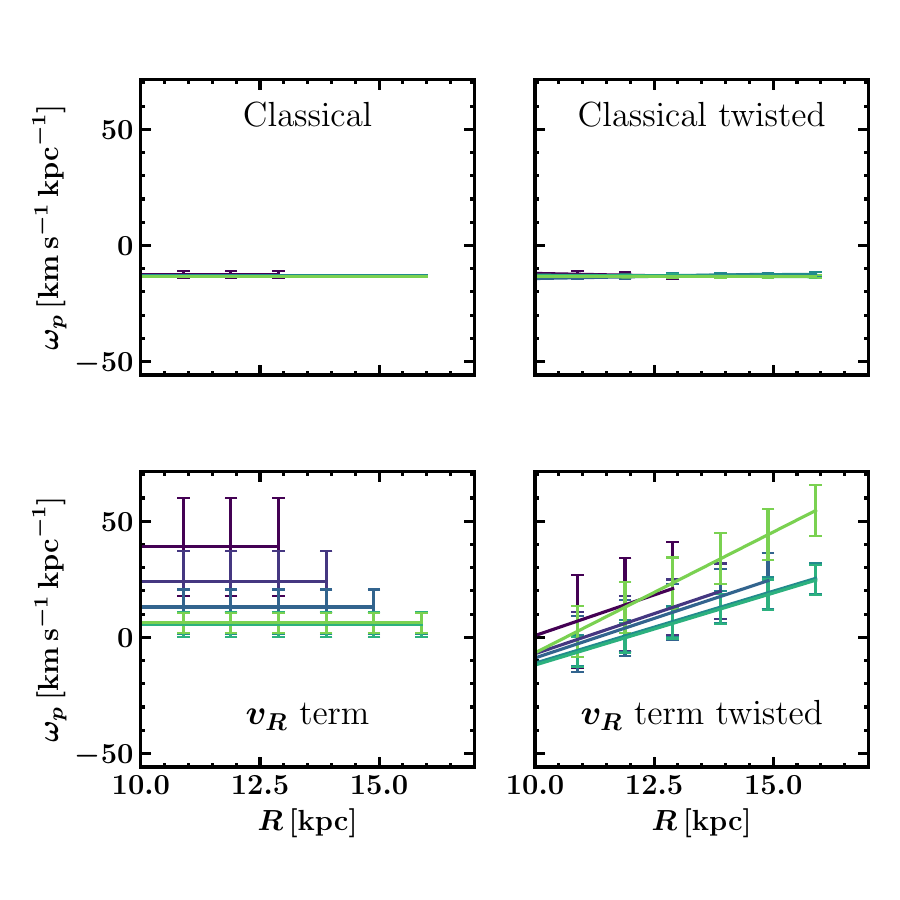}
    \caption{Maximum warp heights ($h$ -- left group of panels), line of nodes ($\phi_w$ -- centre group  of panels) and warp pattern speed ($\omega_p$, right group of panels) as a function of radius for our four models, where the data is truncated at different radii, as indicated.}
     \label{fig:limit}%
 \end{figure*}

Figures~\ref{fig:annuli}~\&~\ref{fig:limit} show the results of our tests with samples selected by Galactocentric radius, either divided into $2\kpc$ wide annuli (Fig.~\ref{fig:annuli}) or with an upper limit on $R$ (Fig.~\ref{fig:limit}). In each case we show $h(R)$, $\phi_w(R)$ and $\omega_w(R)$ for the best fitting model with indicative error bars that are symmetrical around the best fit.  

The maximum warp heights as a function of radius, $h(R)$, are consistent with the values found for our main sample for the two \emph{Classical} models, with a slight tendency in some cases for lower warp heights in the outer parts of the warp. These still appear to be consistent within the uncertainties. The two \emph{$v_R$ term} models are very much consistent for all the samples truncated at different $R$, as seen in Fig.~\ref{fig:limit}, but this is not the case for the samples divided into $2\kpc$ wide annuli, as seen in Fig.~\ref{fig:annuli} (lower left panels).  
In this case the warp heights are significantly higher than for \emph{$v_R$ term} models fit to the other samples -- much closer to the values found for the \emph{Classical} models. It is noticeable that these fits are inconsistent with one another where they overlap, and that the radial gradient of each is negative. Because we have to start this model at a non-zero height 
(to allow for normal behaviour in a warp at these radii), the fitting algorithm is able to find solutions with a significant height but never with a positive gradient in $h(R)$. This allows it to find these solutions without `paying the price' of a positive $h(R)$ gradient elsewhere in the disc that would presumably harm the fit with the data. Note that the \emph{$v_R$ term}  models -- unlike the \emph{Classical} models -- explicitly include the gradient of the warp height with radius in the model (eq.~\ref{eq:cheng_model}). These models are completely unrealistic physically, and should be thought of as a numerical difficulty introduced by using the piecewise linear model for $h(R)$ for these samples.

The lines of nodes for the two \emph{Classical} models applied to the different data samples tell an interesting, hard to disentangle, story which we discuss in Sect.~\ref{sec:line_of_nodes}. The samples divided into $2\kpc$ wide annuli (Fig.~\ref{fig:annuli}) appear to show a leading spiral which joins up neatly and consistently for the \emph{twisted} model. This is the exact opposite of that found with our full sample or when we set a stricter limit of 1000 stars per bin. We can gain a bit more insight by looking at results for the \emph{Classical twisted} model when we truncate the sample at different radii (upper row, fourth panel of Fig.~\ref{fig:limit}, also shown as a polar plot in Fig~\ref{fig:robustLON}). Depending on the cut we take, we can either end up with a leading or trailing spiral of the line of nodes. This does not follow a simple trend, being leading if we truncate at $R=13$ or $14\kpc$, trailing if we truncate at $15$ or $16\kpc$, and leading again with truncations at larger radii.

These results lead us to conclude that our analysis is not capable of providing a robust answer to the question of the twist of the warp's line of nodes.

The results for the pattern speed of the warp for the different data samples are, for the \emph{Classical models}, consistent with our main samples' values, sometimes with large uncertainties. The pattern speed found for $2\kpc$ annuli with the \emph{$v_R$ terms} are, like the results for $h(R)$, closer to those for the full sample with the \emph{Classical} models. This is again due to the numerical difficulty introduced by using the piecewise linear model for $h(R)$ for these samples -- there is a strong correlation between $h(R)$ and $\omega_p$ in the models (eqns.~\ref{eq:standard_model}~\&~\ref{eq:cheng_model}) so a problem with the description of $h(R)$ transfers naturally to a problem with $\omega_p$.

 \section{ADQL query} \label{app:ADQL}

 Our sample is selected from the \href{https://gea.esac.esa.int/archive/}{\Gaia\ archive's} table
 \texttt{gaiadr3.gaia\_source\_lite}, cross-matched with its table \texttt{external.gaiaedr3\_distance} which contains the \cite{bailer-jones_estimating_2021} distance estimates. We use the following ADQL query:
 \begin{verbatim}
     SELECT source_id, ra, dec, 
     parallax, parallax_error, pmra, pmdec, 
     phot_g_mean_mag, bp_rp, ruwe, 
     r_med_photogeo, r_lo_photogeo, r_hi_photogeo, 
     radial_velocity, radial_velocity_error 
     FROM external.gaiaedr3_distance
     JOIN gaiadr3.gaia_source_lite 
     USING (source_id)
     WHERE 0=CONTAINS(POINT('ICRS',ra,dec), 
         CIRCLE('ICRS',81.28,-69.78,12)) 
     AND 0=CONTAINS(POINT('ICRS',ra,dec), 
         CIRCLE('ICRS',13.1583,-72.8003,6)) 
     AND radial_velocity IS NOT NULL
     AND parallax IS NOT NULL
 \end{verbatim}
 We make the final selections to remove stars associated with 47 Tucanae, on parallax uncertainty and RUWE within our data processing.
 
\end{appendix}
\end{document}